\documentclass[12pt]{article}
\usepackage{latexsym}
\usepackage{amsmath}
\renewcommand{\epsilon}{\varepsilon}

\usepackage{amsmath,bm}
\usepackage{amssymb}
\usepackage{xcolor}
\usepackage{amsthm}
\usepackage[ansinew]{inputenc}
\usepackage{graphicx}
\usepackage{epsfig}
\usepackage{dsfont}
\usepackage{bbm}
\usepackage{url}
\usepackage{placeins}
\usepackage[FIGTOPCAP]{subfigure} 
 \usepackage{multirow}

 \usepackage[authoryear]{natbib}
\newtheorem{satz}{Theorem}[section]

\newtheorem{algorithm}[satz]{Algorithm}

\def\3{\ss}

\newcommand{\bea}{\begin{eqnarray*}}
	\newcommand{\eea}{\end{eqnarray*}}
\newcommand{\be}{\begin{eqnarray}}
\newcommand{\ee}{\end{eqnarray}}

\newcommand{\ba}{\begin{array}}
	\newcommand{\ea}{\end{array}}

\def\3{\ss}

\begin{document}
	
	\title{{\bf Survival analysis under non-proportional hazards: investigating non-inferiority or equivalence in time-to-event data }}

	\author{Kathrin M\"ollenhoff \\
	\small Department of Mathematics and Computer Science, \\
	\small	Eindhoven University of Technology, Eindhoven, The Netherlands and \\
	\small University of Cologne, Faculty of Medicine and University Hospital, Cologne, Germany \\
	Achim Tresch \\
		\small University of Cologne, Faculty of Medicine and University Hospital, Cologne, Germany and\\ 
		\small 	CECAD, University of Cologne, Cologne, Germany and \\
		\small Center for Data and Simulation Science, University of Cologne, Germany
	}

	\pdfminorversion=4
	\maketitle
	
	\begin{abstract}

The classical approach to analyze time-to-event data, e.g. in clinical trials, is to fit Kaplan-Meier curves yielding the treatment effect as the hazard ratio between treatment groups. Afterwards commonly a log-rank test is performed in order to investigate whether there is a difference in survival, or, depending on additional covariates, a Cox proportional hazard model is used. 
However, in numerous trials these approaches fail due to the presence of non-proportional hazards, resulting in difficulties of interpreting the hazard ratio and a loss of power. 
When considering equivalence or non-inferiority trials, the commonly performed log-rank based tests are similarly affected by a violation of this assumption.  
Here we propose a parametric framework to assess equivalence or non-inferiority for survival data. 
We derive pointwise confidence bands for both, the hazard ratio and the difference of the survival curves. Further we propose a test procedure addressing non-inferiority and equivalence by directly comparing the survival functions at  certain time points or over an entire range of time.
We demonstrate the validity of the methods by a clinical trial example and by numerous simulation results.

\end{abstract}

\vskip-.2cm
\noindent Keywords and Phrases: equivalence, non-inferiority, survival analysis, time-to-event data, non-proportional hazards, confidence band, hazard ratio, log-rank test

\parindent 0cm

\maketitle

\section{Introduction}
\label{sec1}
\def\theequation{1.\arabic{equation}}
\setcounter{equation}{0}

Time-to-event outcomes are frequently observed in medical research, for instance in the area of oncology or cardiovascular diseases.  
A commonly addressed issue is the comparison of a test to a reference treatment with regard to survival. 
For this purpose an analysis based on Kaplan-Meier curves (see \cite{kaplan1958}), followed by a log-rank test (see, for example, \cite{kalbfleisch1983}) is still the most popular approach. 
Additionally, adjusting for multiple covariates, Cox's proportional hazards model (\cite{cox1972}) has been extensively used in the last decades (for some examples see \cite{cox1984,klein2006} among many others).
In case of addressing non-inferiority or equivalence, extensions of the log-rank test investigating the vertical distance between the survival curves have been proposed by \cite{wellek1993} and \cite{com1993}.
These approaches owe much of their popularity to the fact that they do not rely on assumptions on the distribution of event times.
Moreover, a direct interpretation is obtained by summarizing  the treatment effect in one single parameter, given by the hazard ratio of the two treatments, assumed to be constant over time.

However, this assumption has been heavily criticized (see for example \cite{hernan2010,uno2014}), and is in practice rarely assessed or even obviously violated (see \cite{li2015} and \cite{jachno2019} for an overview on the awareness of this assumption). In particular if short- and long-term benefits of the two treatments differ for instance in situations where a surgical treatment is compared to a non-surgical one (see for example \cite{howard1997} and the references therein), the assumption of proportional hazards is questionable and should be carefully investigated.  
The most obvious sign for a violation of this assumption are survival curves crossing each other. However, often other techniques are required, for example graphical methods 
(see \cite{grambsch1994} for an overview) or statistical tests (see for example \cite{gill1987}).

One of the advantages of the standard methodology based on Kaplan-Meier curves and the log-rank test is that equivalence hypotheses can be formulated using one parameter, that is the hazard ratio. If this ratio changes over time, both, an alternative measure of treatment effect, and a suitable definition of equivalence have to be found. 
For instance, \cite{royston2011} introduce the restricted mean survival time to overcome this issue. 
Moreover, an alternative to commonly used log-rank based tests of equivalence has been proposed by \cite{martinez2017} under the assumption of proportional odds. 
Further these authors show that type I errors for a log-rank based test are higher than the nominal level if the assumption of proportional hazards doesn't hold.
Finally, in a recent paper \cite{shen2020} proposes an alternative test for equivalence based on two other non-parametric models, which can also be used if neither hazards, nor odds, are proportional.

Methods that employ parametric survival models are less common than the above-mentioned semi-or non-parametric methods.
However, a correctly specified parametric survival model provides numerous advantages, as for instance more precise estimates (see, for example, \cite{klein2006}) or the possibility of making predictions. Inference based on parametric models can be very precise even in case of misspecification, as demonstrated by \cite{subramanian2013}, who develop simultaneous confidence bands for parametric survival curves and compare them to non-parametric approaches based on Kaplan-Meier estimates. 

In this paper we develop new methodology in two directions. We address the issue of non-inferiority and equivalence testing in presence of non-proportional hazards by presenting a parametric alternative to the classical methodology, without the assumption of a constant hazard-ratio. 
We first present an approach to derive pointwise confidence bands for the difference of two survival curves and the hazard ratio over time, respectively, by both making inference of the asymptotic distribution of these curves and using a bootstrap approach.
Second we use these confidence intervals to assess equivalence or non-inferiority of the two treatments.
Finally all our methods are illustrated by a clinical trial example and by means of a simulation study.


\section{Methods}
\label{sec2}
\def\theequation{2.\arabic{equation}}
\setcounter{equation}{0}


Consider two samples of size $n_1$ and $n_2$ respectively, resulting in a total sample size of $n=n_1+n_2$. Let $Y_{1,1},\ldots,Y_{1,n_1}$ and $Y_{2,1},\ldots,Y_{1,n_2}$ denote independent random variables representing survival times for individuals allocated to two (treatment) groups, observing a time range given by $\mathcal T=\left[ 0, t_{max} \right]$, where $0$ denotes the start of the observations. Assume that the distribution functions $F_1$ and $F_2$ of $Y_{1,j},\ j=1,\ldots,n_1$ and $Y_{2,j}\ j=1,\ldots,n_2$, respectively, are absolutely continuous with densities $f_1$ resp. $f_2$. Consequently the probability of experiencing an event for an individual $j$ of the $\ell$-th sample before time $t$ can be written as $$F_\ell(t)=\mathbb{P}(Y_{\ell,j}< t)=\int_{0}^{t}f_{\ell}(u) du,\ \ell=1,2.$$ 
Further denote the corresponding survival functions by $S_\ell:=1-F_\ell$ and the hazard rates by $h_\ell:=\tfrac{f_\ell}{S_\ell}$. The cumulative hazard function is given by $H_\ell(t)=-\log(S(t))$, $\ell=1,2$.

For the sake of simplicity we do not assume additional covariates. Further we assume all observations to be randomly right-censored and denote the censoring times of the two samples by $C_{1,1},\ldots,C_{1,n_1}$ and $C_{2,1},\ldots,C_{2,n_2}$ and the corresponding distribution functions by $G_1$ and $G_2$ respectively. Note that these distributions can differ from each other and are assumed to be independent from the $Y_{\ell,j},\ \ell=1,2,\ j=1,\ldots n_\ell$.
We define $\Delta_{\ell,j}=I\{Y_{\ell,j} \geq C_{\ell,j}\},$
indicating whether an individual is censored ($\Delta_{\ell,j}=0$) or experiences an event ($\Delta_{\ell,j}=1$), where $I$ denotes the indicator function. Consequently the observed data $(t_{\ell,j},\delta_{\ell,j})$ is a realization of the bivariate random variable $(T_{\ell,j},\Delta_{\ell,j})$, where $T_{\ell,j}=\min(Y_{\ell,j},C_{\ell,j}),\ \ell=1,2,\ j=1,\ldots n_\ell$. 
In order to make inference on the underlying distributions we consider the likelihood function for group $\ell$ given by
\be\label{mle}
L_\ell(F_\ell,G_\ell)=
\prod_{j=1}^{n_\ell}\left\lbrace f_\ell(t_{\ell,j})^{\delta_{\ell,j}}(1-F_\ell(t_{\ell,j}))^{1-\delta_{\ell,j}}\right\rbrace \cdot
\prod_{j=1}^{n_\ell}\left\lbrace (1-G_\ell(t_{\ell,j}))^{\delta_{\ell,j}}(g_\ell(t_{\ell,j}))^{1-\delta_{\ell,j}}\right\rbrace
,\ee
as censoring times and survival times are assumed to be independent. Hence we can obtain estimates for the densities $f_\ell(t)=f_\ell(\theta_\ell,t)$ and $g_\ell(t)=g_\ell(\psi_\ell,t)$ by deriving the parameters $\hat\theta_\ell$ and $\hat\psi_\ell$ maximizing $\log{L_\ell}$, $\ell=1,2$. Note that if one is not interested in estimating the underlying distribution of the censoring times and $\theta_\ell$ and $\psi_\ell$ have no common parameters, this optimization procedure can be further simplified by just considering the first part in \eqref{mle}, resulting in an objective function given by
\be\label{mle2}
\tilde L_\ell(\theta_\ell)=
\prod_{j=1}^{n_\ell}\left\lbrace f_\ell(\theta_\ell,t_{\ell,j})^{\delta_{\ell,j}}(1-F_\ell(\theta_\ell,t_{\ell,j}))^{1-\delta_{\ell,j}}\right\rbrace,\ \ell=1,2.
\ee

\subsection{Confidence bands}

In the following we will construct pointwise confidence bands for the difference of the survival functions and for the hazard ratio. First we derive an asymptotic approach using the Delta-method  (see \cite{oehlert1992}) and second, we propose an alternative based on a bootstrap procedure, where the latter can also be used when samples are very small or, moreover, if  asymptotic inference cannot be made due to the lack of concrete expressions for the variance. 
In order to simplify calculations, we will consider the log-ratio and therefore the two measures of interest are given by
  \be
  \Delta(t,\theta_1,\theta_2):=S_1(t,\theta_1)-S_2(t,\theta_2)\text{ and } r(t,\theta_1,\theta_2):=\log{\tfrac{h_1(t,\theta_1)}{h_2(t,\theta_2)}}.
  \ee
 Under certain regularity conditions (see \cite{bradley1962}) the Maximum Likelihood estimates (MLE) obtained by maximizing \eqref{mle} or \eqref{mle2} are asymptotically normally distributed, that is
$$
\sqrt{n_\ell}\big(\hat\theta_\ell-\theta_\ell\big)\stackrel{\cal D}{\longrightarrow} \mathcal{N}(0, \mathcal{I}_{\theta_\ell}^{-1}),\ \ell=1,2,
$$
where $\mathcal I_\theta^{-1}$ denotes the inverse of the Fisher information matrix.
This result can be used to make inference about the asymptotic distribution of the estimated survival curves. More precisely, using the Delta-method we obtain for every $t>0$ 
$$
\sqrt{n_\ell}\big(S_\ell(t,\hat\theta_\ell)-S_\ell(t,\theta_\ell)\big)\stackrel{\cal D}{\longrightarrow} \mathcal{N}(0, \tfrac{\partial}{\partial \theta_\ell} S_\ell(t,\theta_\ell)^T \mathcal I_{\theta_\ell}^{-1}  \tfrac{\partial}{\partial \theta_\ell} S_\ell(t,\theta_\ell)),\ \ell=1,2.
$$
Consequently an estimate of the asymptotic variance of $\Delta(t,\theta_1,\theta_2)$ is given by
\begin{align}\label{var2}
\hat\sigma_\Delta^2:=Var{(\Delta(t,\hat\theta_1,\hat\theta_2))}=\tfrac{1}{n_1}\tfrac{\partial}{\partial \theta_1}S_1(t,\hat\theta_1)^T I_{\hat\theta_1}^{-1} \tfrac{\partial}{\partial\theta_1}S_1(t,\hat\theta_1)+\tfrac{1}{n_2}\tfrac{\partial}{\partial \theta_2}S_2(t,\hat\theta_2)^T I_{\hat\theta_2}^{-1} \tfrac{\partial}{\partial \theta_2}S_2(t,\hat\theta_2),
\end{align}
where $I_{\hat\theta_\ell}$ denotes the observed information matrix, $\ell=1,2$. For sufficiently large samples this asymptotic result can be used to construct  pointwise lower and upper $(1-\alpha)$-confidence bounds, respectively, by
\be\label{conf_sf} L_\Delta(t,\hat\theta_1,\hat \theta_2):=\Delta(t,\hat\theta_1,\hat\theta_2)-z_{1-\alpha} \hat\sigma_\Delta
\text{ and }
U_\Delta(t,\hat\theta_1,\hat \theta_2):=\Delta(t,\hat\theta_1,\hat\theta_2)+z_{1-\alpha} \hat\sigma_\Delta,
\ee
where $z_{1-\alpha}$ denotes the $(1-\alpha)$-quantile of the standard normal distribution. 
More precisely, if $L(t)$ and $U(t)$ denote the $(1-\alpha)$ pointwise lower and the $(1-\alpha )$ pointwise upper confidence bound, respectively, it holds
\begin{align}
\label{conf_bands}
\lim_{n_1,n_2\rightarrow \infty}\mathbb{P}\big(L_\Delta(t,\hat\theta_1,\hat \theta_2)\leq \Delta(t,\theta_1,\theta_2)\big)\geq 1-\alpha\text{ , }
\lim_{n_1,n_2\rightarrow \infty}\mathbb{P}\big(\Delta(t,\theta_1,\theta_2)\leq U_\Delta(t,\hat\theta_1,\hat \theta_2)\big)\geq 1-\alpha
\end{align}
for all $t>0$, where $\alpha$ denotes the prespecified significance level.
The construction of pointwise confidence bands for the log hazard ratio works similarly and the
 asymptotic variance of $r(t)$ is given by
\begin{align}\label{var}
\hat\sigma_r^2:&=Var{(r(t,\hat\theta_1,\hat\theta_2))}=\tfrac{1}{n_1}\cdot\tfrac{\partial}{\partial \theta_1}\log{h_1(t,\hat\theta_1)}^T I_{\theta_1}^{-1} \tfrac{\partial}{\partial\theta_1}\log{h_1(t,\hat\theta_1)}\nonumber\\ &+
\tfrac{1}{n_2}\cdot\tfrac{\partial}{\partial \theta_2}\log{h_2(t,\hat\theta_2)}^T I_{\theta_2}^{-1} \tfrac{\partial}{\partial \theta_2} \log{h_2(t,\hat\theta_2)}.
\end{align}
Consequently $L_d$ and $U_d$ are given by
\be\label{conf_hr} L_d(t,\hat\theta_1,\hat \theta_2):=r(t,\hat\theta_1,\hat \theta_2)-z_{1-\alpha} \hat\sigma_r
\text{, }
U_d(t,\hat\theta_1,\hat \theta_2):=r(t,\hat\theta_1,\hat \theta_2) +z_{1-\alpha}\hat\sigma_r
\ee
and it holds for all $t>0$ that
\begin{align*}
\lim_{n_1,n_2\rightarrow \infty}\mathbb{P}\big(L_d(t,\hat\theta_1,\hat \theta_2)\leq r(t,\theta_1,\theta_2)\big)\geq 1-\alpha\text{ and }
\lim_{n_1,n_2\rightarrow \infty}\mathbb{P}\big(r(t,\theta_1,\theta_2)\leq  U_d(t,\hat\theta_1,\hat \theta_2)\big)\geq 1-\alpha.
\end{align*}
A concrete numerical example for calculating the variances in \eqref{var2} and \eqref{var} and the corresponding confidence bounds assuming a Weibull distribution are deferred to Section 2 in the Supplemental Material and the finite sample properties for to this scenario are presented in Section \ref{sec:sim}.

If sample sizes are rather small or the variability in the data is high, we propose to obtain estimates for the variances $\hat\sigma_{\Delta}^2$ and $\hat\sigma_r^2$ by using a bootstrap approach, taking the right-censoring into account. This approach has the advantage that it can also be used if a formula for the asymptotic variance is not obtainable, for instance due to numerical difficulties. The following algorithm explains the procedure for $\Delta(t_0,\theta_1,\theta_2)$ and it can directly be adapted to $r(t_0,\theta_1,\theta_2)$.
\begin{algorithm}{(Parametric) Bootstrap Confidence Intervals for $\Delta(t_0,\theta_1,\theta_2)$.}
	
\begin{itemize}
	\item[1.] Calculate the MLE  $\hat\theta_\ell$ and $\hat\psi_\ell,\ \ell=1,2$, from the data by maximizing \eqref{mle}.
	\item[2(a).] Generate survival times $y_{\ell,1}^*,\ldots,y_{\ell,n_\ell}^*$  from $F_\ell(\hat\theta_\ell)$, $\ell=1,2$. 
	Further generate the corresponding censoring times $c_{\ell,1}^*,\ldots,c_{\ell,n_\ell}^*$ by sampling from the distributions $G_\ell(\hat\psi_\ell)$, $\ell=1,2$. 
	If $y_{\ell,j}^*>c_{\ell,j}^*$, the observation is censored (i.e. $\delta_{\ell,j}^*=0$), $j=1,\ldots,n_\ell$.  
	The observed data is given by $(t_{\ell,j}^*,\delta_{\ell,j}^*)$, $t_{\ell,j}^*=min(y_{\ell,j}^*,c_{\ell,j}^*)$, $j=1,\ldots,n_\ell$, $\ell=1,2$.
	\item[2(b).] Calculate MLE $\hat\theta_\ell^*$ for the bootstrap sample from the $t_{\ell,j}^*,\  j=1,\ldots,n_\ell,\ \ell=1,2$, and calculate the difference of the corresponding survival functions at $t_0$, that is 
	\be
	\Delta^*:=\Delta(t_0,\hat\theta_1^*,\hat\theta_2^*)=S_1(t_0,\hat\theta_1^*)-S_2(t_0,\hat\theta_2^*).
	\ee
		\item[3.] Repeat steps $2(a)$ and $2(b)$ $n_{boot}$ times, yielding $\Delta^*_1,\ldots, \Delta^*_{n_{boot}}$ and calculate an estimate for the variance $\hat\sigma_{\Delta}^2$ by
		\be\label{bootstrap_var} \hat\sigma_{\Delta}^2=\frac{1}{n_{boot}-1}\sum_{k=1}^{n_{boot}}(\Delta^*_k-\bar\Delta^*)^2, \ee
		where $\bar\Delta^*$ denotes the mean of the $\Delta_i^*,\ i=1,\ldots,n_{boot}$.
\end{itemize}	
\label{alg1}\end{algorithm}
Finally the estimate $ \hat\sigma_{\Delta}^2$ in \eqref{bootstrap_var} is used to calculate the confidence band in \eqref{conf_sf}.
Note that the procedure described in Algorithm \ref{alg1} is a parametric bootstrap as it is based on estimating the parameters $\hat\theta_\ell,\ \hat\psi_\ell,\ \ell=1,2$. A non-parametric alternative, given by resampling the observations (see \cite{efron1994}), could also be implemented. 

\subsection{Equivalence and non-inferiority tests}

We are aiming to compare the survival functions of two (treatment) groups which is commonly addressed by testing the null hypothesis that the two survival functions are identical versus the alternative that they are differing at any time point (for an overview, see, for example, \cite{klein2006}). More precisely the classical hypotheses are given by
$$
H_0: S_1(t,\theta_1)=S_2(t,\theta_2) \text{ for all $t\in\cal T$ against } H_1:S_1(t,\theta_1)\neq S_2(t,\theta_2)\text{ for a $t\in\cal T$}.
$$
In some situations one might be more interested in observing non-inferiority of one treatment among another or equivalence of the two treatments, meaning that we allow a deviation of the survival curves of a prespecified threshold instead of testing for equality. This can be done for a particular point in time, say $t_0$, or over an entire interval, say $[t_1,t_2]$.  We now consider the difference in survival at a particular point in time $t_0$. The corresponding hypotheses are then given by
\be\label{hypotheses_ni}
H_0: S_1(t_0,\theta_1)-S_2(t_0,\theta_2)\geq \delta \text{ against } H_1 :S_1(t_0,\theta_1)-S_2(t_0,\theta_2)< \delta
\ee
for a non-inferiority trial observing whether a test treatment is non-inferior to the reference treatment (which is stated in the alternative hypothesis). Considering equivalence, the hypotheses are given by
\be\label{hypotheses_eq}
H_0: \left| S_1(\theta_1,t_0)-S_2(\theta_2,t_0)\right| \geq \delta \text{ against } H_1 :  \left|S_1(\theta_1,t_0)-S_2(\theta_2,t_0)\right|< \delta.
\ee
The same questions can be addressed considering the (log) hazard ratio, resulting in
the hypotheses analogue to \eqref{hypotheses_ni} given by
\be\label{hypotheses_ni_2}
H_0:  \log{\tfrac{h_1(\theta_1,t_0)}{h_2(\theta_2,t_0)}}\geq \epsilon \text{ against } H_1 :   \log{\tfrac{h_1(\theta_1,t_0)}{h_2(\theta_2,t_0)}}< \epsilon
\ee
for a non-inferiority trial and 
\be\label{hypotheses_eq_2}
H_0: \left| \log{\tfrac{h_1(\theta_1,t_0)}{h_2(\theta_2,t_0)}}\right| \geq \epsilon \text{ against } H_1 :  \left| \log{\tfrac{h_1(\theta_1,t_0)}{h_2(\theta_2,t_0)}}\right|< \epsilon
\ee
for addressing equivalence. 
The choice of the margins $\delta>0$ and $\epsilon>0$ has to be verified in advance with great care combining statical and clinical expertise. From a regulatory point of view there is no fixed rule but general advice can be found in a guideline of the European Medicines Agency \cite{ema2004}. Following recent literature margins $\delta$ for the survival difference are frequently chosen between $0.1$ and $0.2$, for detailed discussions on that topic we refer to \cite{d2003,da2009,wellek2010testing}, who also investigate the correspondence between $\delta$ and the hazard ratio threshold $\epsilon$.

We now use the confidence bounds derived in \eqref{conf_sf} to construct an asymptotic $\alpha$-level test for \eqref{hypotheses_ni}. More precisely, the null hypothesis concerning non-inferiority in  \eqref{hypotheses_ni} is rejected if the upper bound of the confidence interval is below the equivalence margin, that is
\be\label{test_diff_ni}U_\Delta(t_0,\hat\theta_1,\hat\theta_2)\leq \delta.
\ee
Further, rejecting $H_0$ whenever 
\be\label{test_diff_eq}U_\Delta(t_0,\hat\theta_1,\hat\theta_2)\leq \delta\text{ and } L_\Delta(t_0,\hat\theta_1,\hat\theta_2)\geq -\delta
\ee  
yields an  equivalence test for the hypotheses in  \eqref{hypotheses_eq}. According to the intersection-union-principle\cite{berger1982}, the same $(1-\alpha)$-confidence bounds $L_\Delta(t_0,\hat\theta_1,\hat\theta_2)$ and $U_\Delta(t_0,\hat\theta_1,\hat\theta_2)$ are used for both, the non-inferiority and the equivalence test. The following theorem states that this yields an asymptotic $\alpha$-level test.

\begin{satz}
The test described in \eqref{test_diff_eq} yields an asymptotic $\alpha$-level equivalence tests for the hypotheses in \eqref{hypotheses_eq}. More precisely it holds for all $t_0\in\cal T$
$$
\lim_{n_1,n_2\rightarrow\infty}\mathbb{P}_{H_0}(U_\Delta(t_0,\hat\theta_1,\hat\theta_2)\leq \delta, L_\Delta(t_0,\hat\theta_1,\hat\theta_2)\geq -\delta)\leq \alpha.
$$
\end{satz}

The proof is left to Section 1 in the Supplemental Material and a similar procedure can be applied for testing the hypotheses in \eqref{hypotheses_ni_2} and \eqref{hypotheses_eq_2}.

\subsubsection{Comparing survival over entire time intervals}\label{ssec:interval}
In some situations it might be interesting to compare survival not only at one particular point in time but over an entire period of time $[t_1,t_2]$. 
For instance, extending the equivalence test stated in \eqref{hypotheses_eq} to this situation yields the hypotheses
\be\label{hypotheses_eq2}
H_0: \max_{t\in [t_1,t_2]}\left| S_1(t,\theta_1)-S_2(t,\theta_2)\right| \geq \delta \text{ against } H_1 : \max_{t\in [t_1,t_2]} \left|S_1(t,\theta_1)-S_2(t,\theta_2)\right|< \delta
\ee
and a similar extension can be formulated for the non-inferiority test \eqref{hypotheses_ni} and the tests on the hazard ratio \eqref{hypotheses_ni_2} and \eqref{hypotheses_eq_2}, respectively.  
In this case we reject the null hypothesis in \eqref{hypotheses_eq} if, for all $t$ in $[t_1,t_2]$, the confidence bounds $L_\Delta(t)$ and $U_\Delta(t)$ derived in \eqref{conf_sf} are included in the equivalence region $\left[ -\delta,\delta\right] $.

\section{Finite sample properties}
\label{sec:sim}
\def\theequation{3.\arabic{equation}}

In the following we will investigate the finite sample properties of the proposed methods by means of a simulation study. 
We consider two scenarios corresponding to the choice of a Weibull distribution of survival times. In a third case we investigate the robustness of our approach by generating data according to a log-logistic distribution but still fitting a Weibull model. 
We assume (randomly) right-censored observations in combination with an administrative censoring time in both scenarios meaning a fixed time point of last follow-up which we denote by $t_{max}$. All results are obtained by running $n_{sim}=1000$ simulations and $n_{boot}=500$ bootstrap repetitions.

For all three scenarios we will calculate confidence bands for both the difference of the survival curves and the log hazard ratio and observe their coverage probabilities. 
For the difference of the survival curves, we will investigate the tests on non-inferiority and equivalence proposed in \eqref{test_diff_ni} and \eqref{test_diff_eq}, respectively. For this purpose we will vary both, the particular time point under consideration $t_0$ and the non-inferiority/equivalence margin $\delta$. 
More precisely we will consider three different choices for this margin ranging from rather conservative to liberal, namely $\delta=0.1,0.15$ and $0.2$.
The test concerning the hypotheses formulated in terms of the log hazard ratio (see \eqref{hypotheses_ni_2} and \eqref{hypotheses_eq2}) is conducted in the same manner, results are omitted here.
In addition to our new method we also evaluate all scenarios using a non-parametric approach as described in Section 5.2. of \cite{com1993}. More precisely we construct confidence bounds for the difference of two Kaplan-Meier curves by estimating the variance using the formula by \cite{greenwoodnatural}. 
Due to the sake of brevity the latter results will be deferred to Section 3 of the Supplemental Material.

For the first two scenarios we assume the data in both treatment groups to follow a Weibull distribution, that is $F_\ell(t,\theta_\ell)=1-\exp{\left\lbrace -(t/\theta_{\ell,2})^{\theta_{\ell,1}}\right\rbrace }$, $\ell=1,2$, where $\theta_{\ell,1}$ denotes the shape parameter and $\theta_{\ell,2}$ the scale parameter corresponding to treatment group $\ell=1,2$. We consider a time range (in months) given by $\mathcal T=[0,9]$, where $t_{max}=9$ is the latest point of follow up. 

\begin{figure}
	\scriptsize
	\begin{center}
		(a)\subfigure{\includegraphics[width=0.3\textwidth]{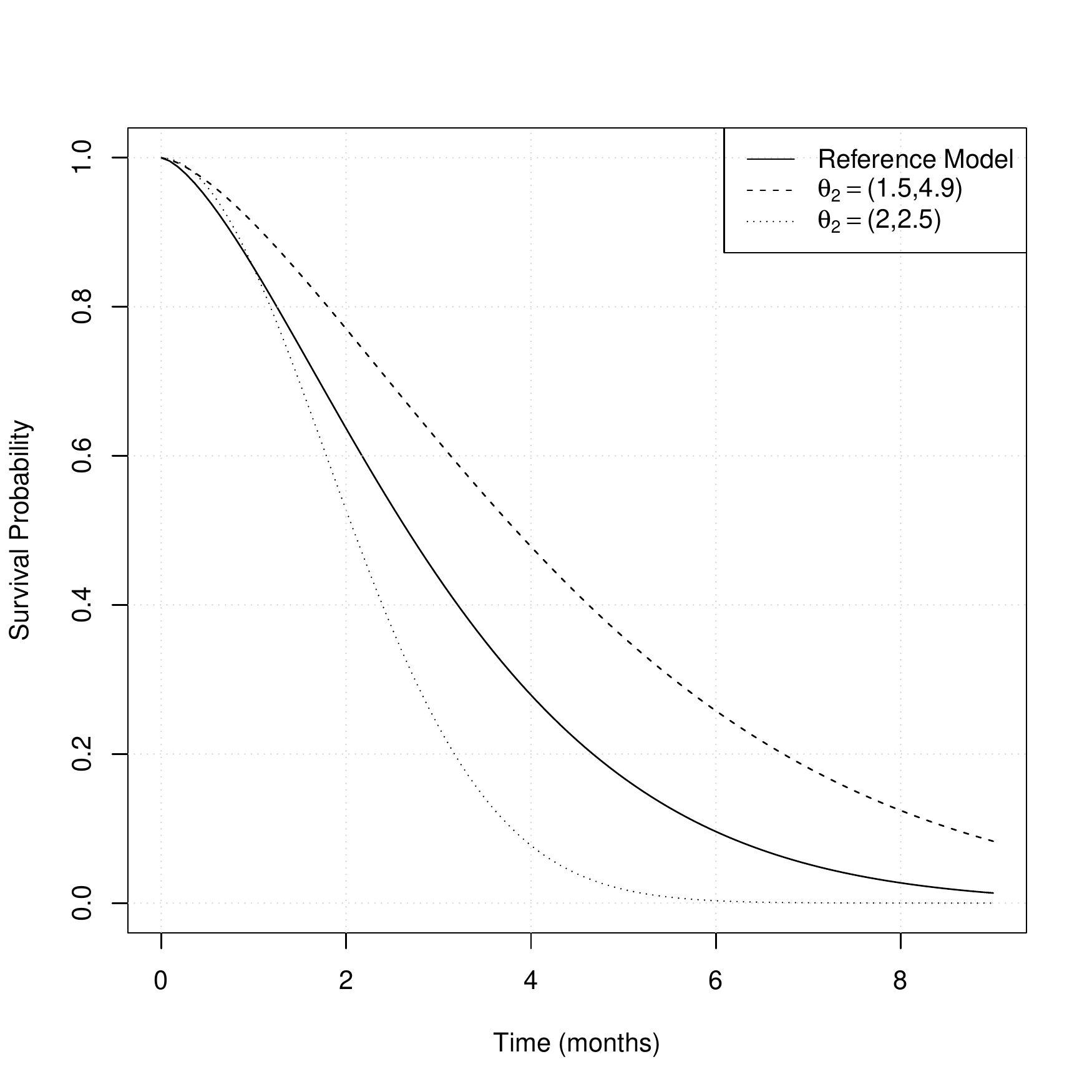}}
		(b)\subfigure{\includegraphics[width=0.3\textwidth]{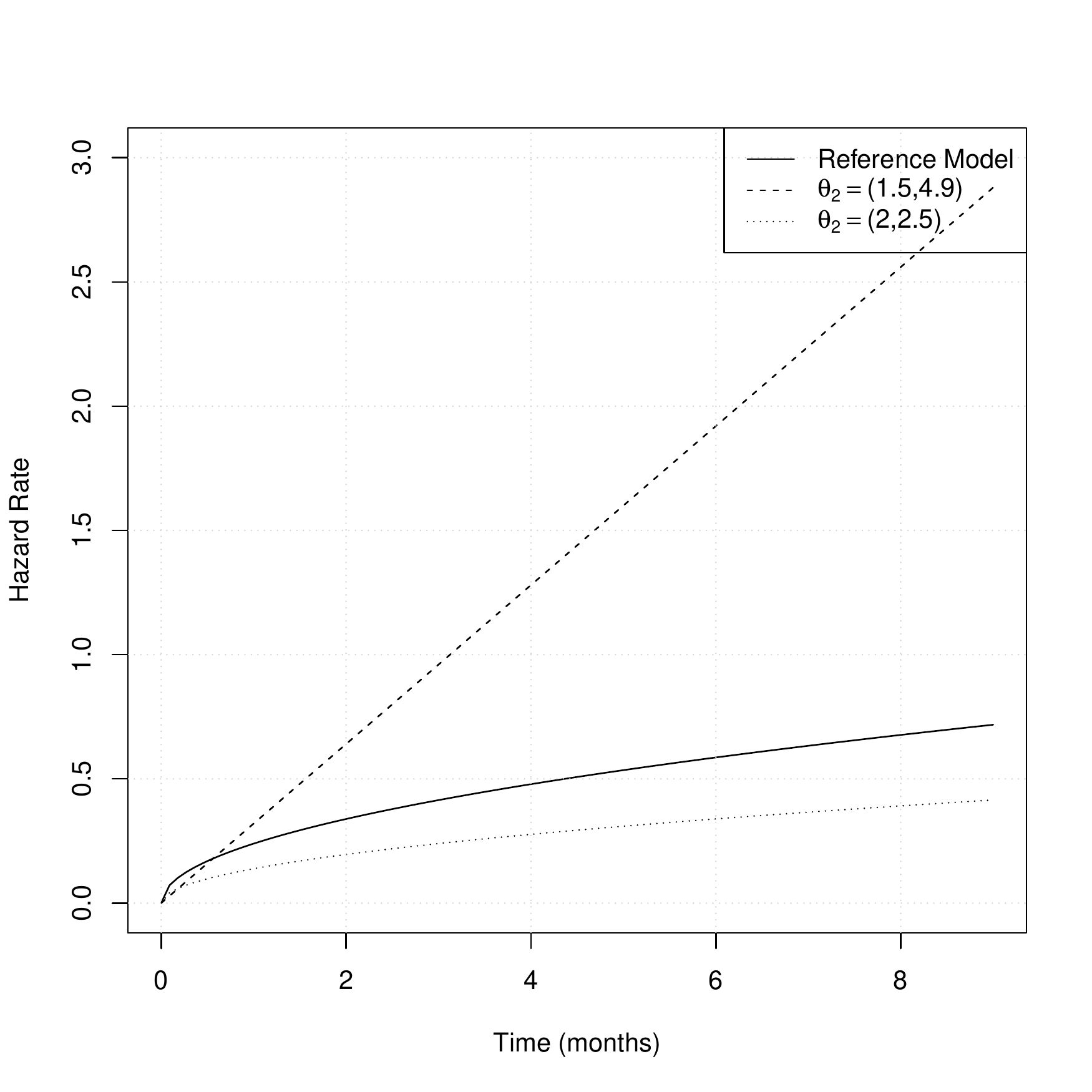}}
		(c)\subfigure{\includegraphics[width=0.3\textwidth]{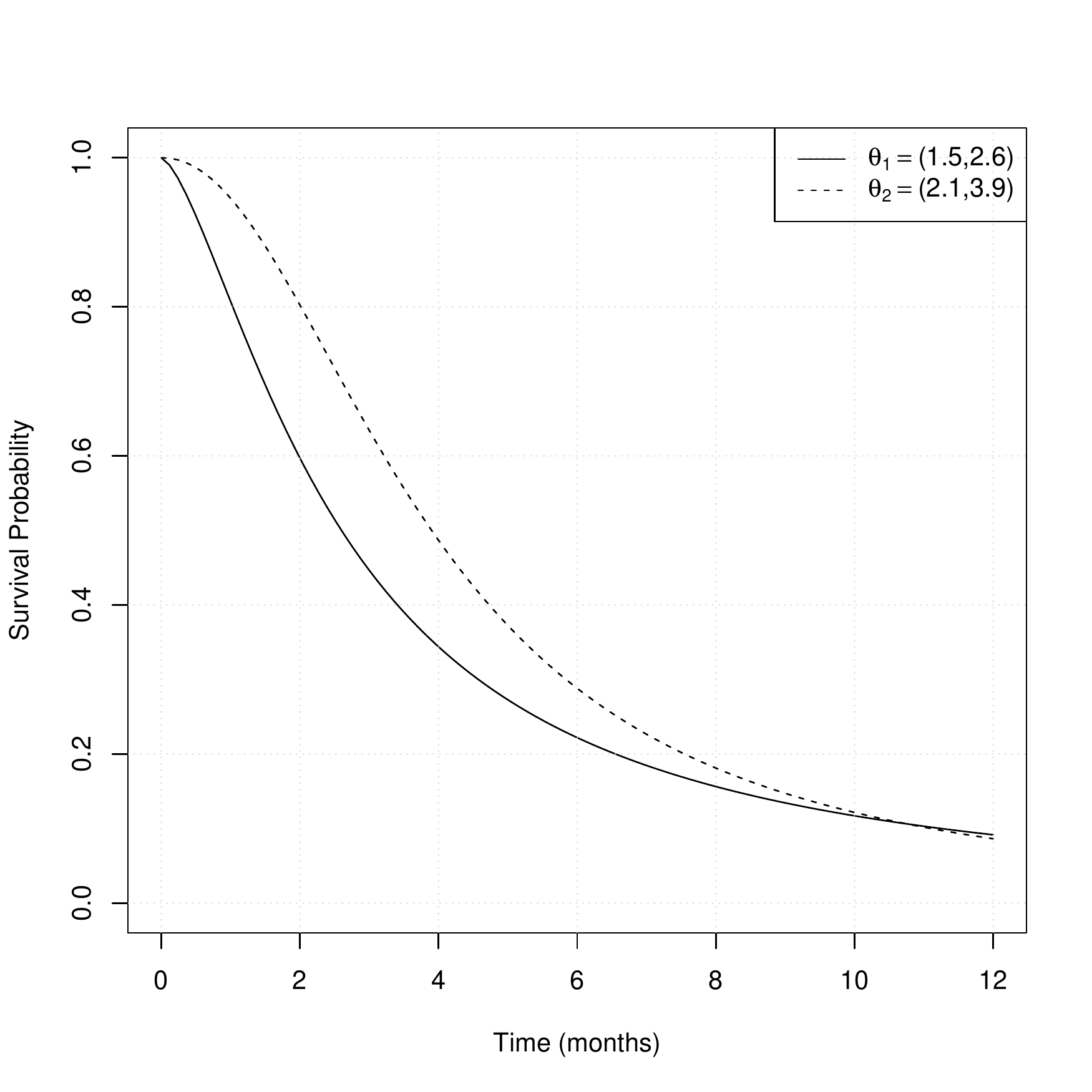}}
		\caption{\small The three scenarios under consideration used for simulating type I error rates and coverage probabilites. (a): Survival curves for Scenarios \eqref{scen1a} and \eqref{scen1b} with $ \theta_1=(1.5,3.4)$, $\theta_2\in\left\lbrace (1.5,4.9),(2,2.5)\right\rbrace $.  (b): Corresponding hazard rates. (c):  Survival curves for Scenario \eqref{scen2}. }\label{fig1}\end{center}
	\end{figure}

For the first configuration we choose 
\be\label{scen1a}
\theta_1=(1.5,3.4),\ \theta_2^1=(1.5,4.9),\ \theta_2^2=(1.5,3.7),
\ee
where $\theta_1$ corresponds to the reference model and the second model is varied by its scale parameter. Here $ \theta_2^1$ is used for investigating the type I errors and coverage probabilities and $ \theta_2^2$ for simulating the power, respectively. As an example, Figure \ref{fig1}(a) displays the survival curves for a choice of $ \theta_2^1$. Both configurations result in a constant log hazard ratio of $\log{(1.5)}\approx 0.4$, representing the situation of proportional hazards.
We assume the censoring times to be exponential distributed and choose the rates of the two groups such that a censoring rate of approximately $25\%$ results, that is a rate of $\psi_1=0.1$ for the reference model and rates of $\psi_2^1=0.09$ and $\psi_2^2=0.05$ for $ \theta_2^1$ and $ \theta_2^2$, respectively. 



In order to investigate the effect of  non-proportional hazards we consider a second scenario of intersecting survival curves, where we keep the reference model specified by $\theta_1$ and all other configurations as above, but vary the parameters of the second model, resulting in
     \be\label{scen1b}
    \theta_1=(1.5,3.4),\ \theta_2^1=(2,2.5),\ \theta_2^2=(2,3.4).
     \ee        
Here the choice of $ \theta_2^1$ is used for investigating the type I errors and coverage probabilities and  $\theta_2^2$ for simulating the power, respectively. 
Again, we consider censoring rates of approximately $25\%$ for both treatment groups, meaning a rate of $\psi_1^1=0.14$ and $\psi_2^1=0.1$ for $ \theta_2^1$ and $ \theta_2^2$, respectively.

\begin{figure}
	\scriptsize
	\begin{center}
		(a)\subfigure{\includegraphics[width=0.46\textwidth]{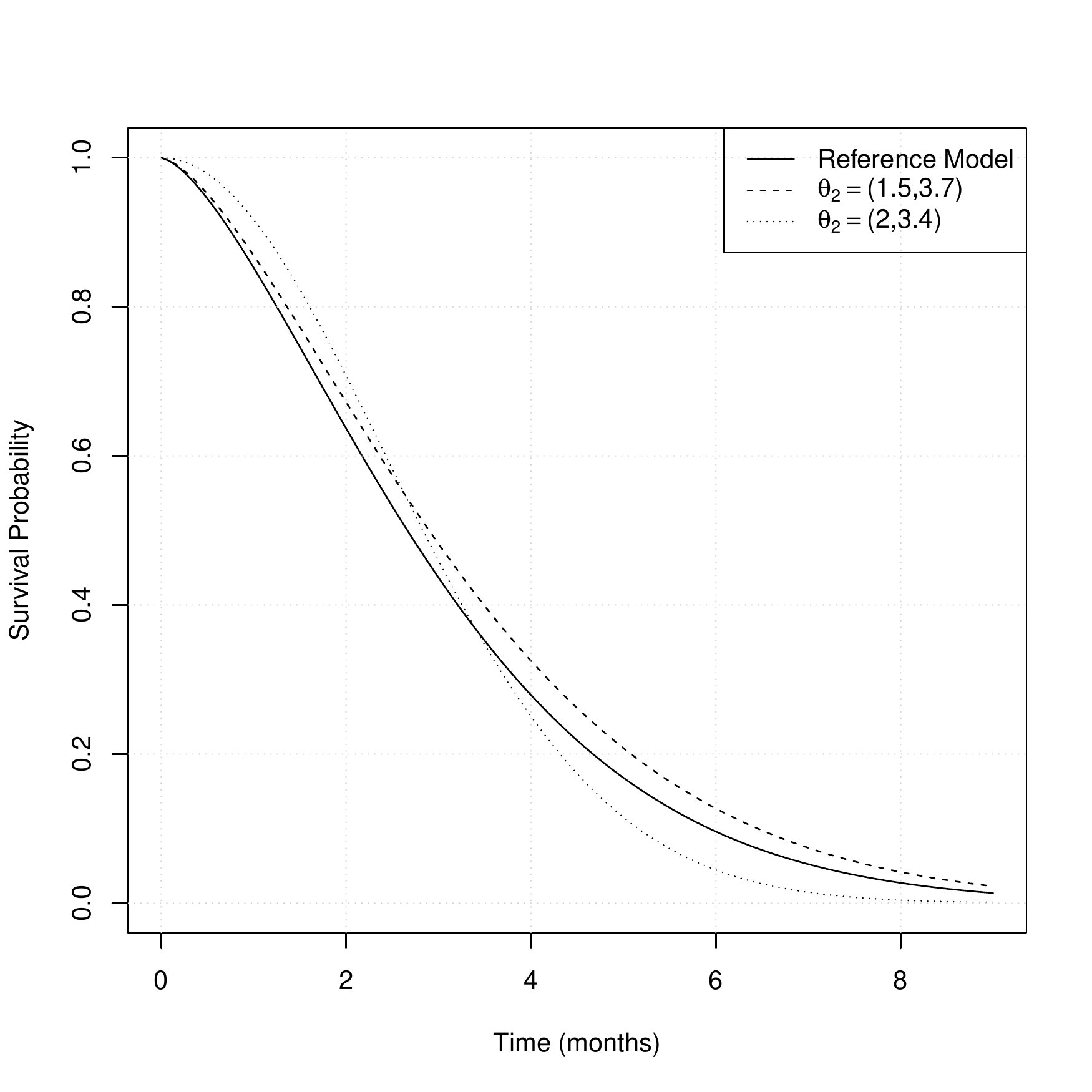}}
		(b)\subfigure{\includegraphics[width=0.46\textwidth]{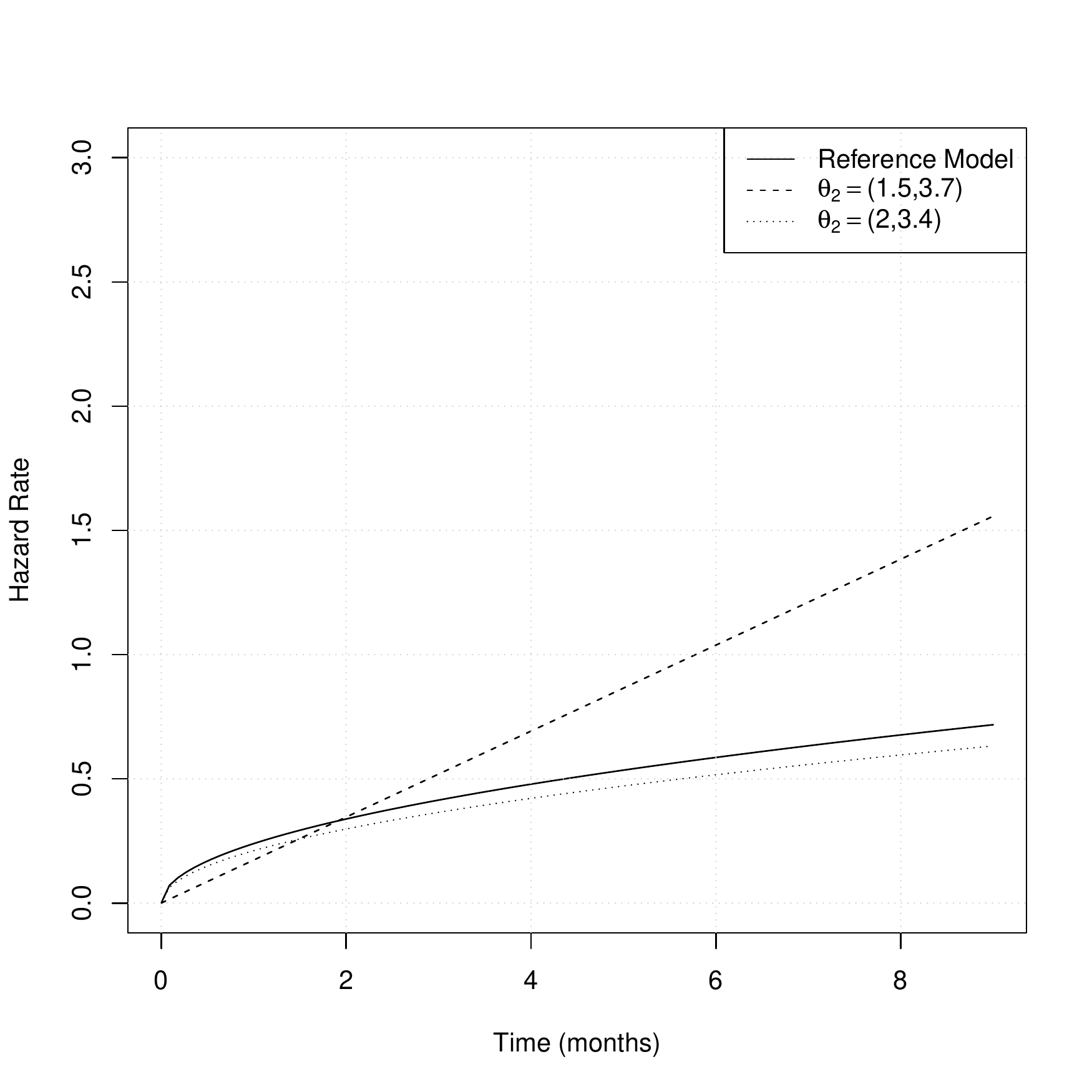}}
		\caption{\small The two scenarios used for simulating the power. (a): Survival curves for Scenarios \eqref{scen1a} and \eqref{scen1b} with $ \theta_1=(1.5,3.4)$, $\theta_2\in\left\lbrace (1.5,3.7),(2,3.4)\right\rbrace $.  (b): Corresponding hazard rates.}\label{fig1b}\end{center}
	\end{figure}

For the third scenario we generate the survival times according to a log-logistic distribution. More precisely we choose $F_\ell(t,\theta_\ell)=1-\tfrac{1}{1+(t/\theta_{\ell,1})^{-\theta_{\ell,2}}}$, $\ell=1,2$. 
We now assume the censoring times to be uniformly distributed on an interval $[0,c_\ell]$, where $c_\ell$ is chosen such that a censoring rate of approximately $20\%$ results, $\ell=1,2$.
We consider a time range (in months) given by $\mathcal T=[0,12]$ and define the scenario by the set of parameters given by
\be\label{scen2}
\theta_1=(1.5,2.6),\ \theta_2=(2.1,3.9),
\ee        
where $\theta_1$ corresponds to the reference model, see Figure \ref{fig1}(c). 

\subsection{Coverage probabilities}

In order to investigate the performance of the confidence bands derived in \eqref{conf_sf} and \eqref{conf_hr} we consider the scenarios described above for three different sample sizes, that is $(n_1,n_2)=(20,20)$, $(n_1,n_2)=(50,50)$ and $(n_1,n_2)=(100,100)$, resulting in total sample sizes given by $n=40,\ 100$ and $200$, respectively. 
We choose a nominal level of $\alpha=0.05$ and calculate both, the asymptotic (two-sided) confidence bands obtained by using the Delta method, and the bands based on the bootstrap described in Algorithm \ref{alg1}, which we call in the following asymptotic bands and bootstrap bands, respectively. 
Confidence bands were constructed for an equidistant grid of $23$ time points ranging from $1.5$ to $6$ months for the first scenario and for a grid of $14$ different points ranging from $1.5$ to $4$ months for the second one, respectively. For the investigations on the situation of misspecification we consider $21$ different time points, ranging from $1$ to $5$ months. 
For each of these time points we check for all approaches whether or not the true value is contained in the calculated confidence interval, which finally yields the simulated coverage probabilities after all simulation runs have been conducted.

\begin{figure}[]
	\begin{center}
		\centering
	\includegraphics[width=0.98\textwidth]{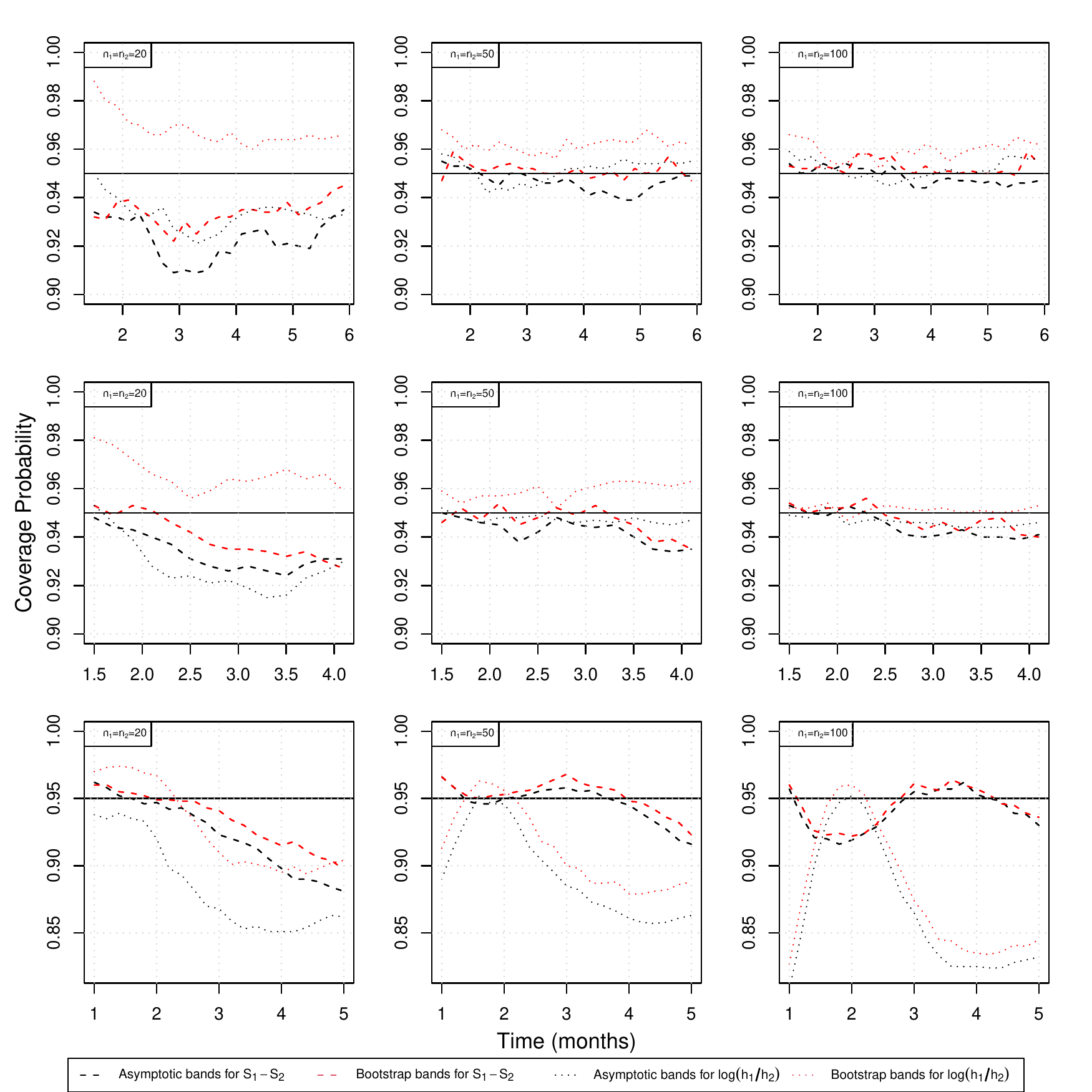}
		\caption{\small Simulated coverage probabilities for scenarios \eqref{scen1a} (first row), \eqref{scen1b} (second row) and the scenario of misspecification \eqref{scen2} (third row) at different time points for sample sizes of $n_1=n_2=20,\ 50,\ 100$ (left, middle, right column). The dashed lines correspond to the confidence bands for the difference of the survival functions \eqref{conf_sf}, the dotted lines to the confidence bands for the log hazard ratio \eqref{conf_hr}. Black lines display the asymptotic bands, red lines the confidence bands based on bootstrap, respectively.}\label{fig2}\end{center}
	\end{figure}

We first consider the two correctly specified scenarios (event distribution of reference and test group is Weibull and modelled as such). 
For the first configuration the hazard ratio is constant over time, for the second it varies between $0.5$ and $2$ on the grid described above, or, equivalently, from $-0.6$ to $0.7$ considering the log hazard ratio.
The first two rows of Figure \ref{fig2} summarize the simulated coverage probabilities for these scenarios. 
In general it turns out that for both scenarios and all approaches, i.e. the asymptotic bands and the bootstrap bands for $S_1-S_2$ and $\log{\tfrac{h_1}{h_2}}$, respectively, the approximation is very precise when sample sizes increase, as the coverage probabilities are very close to the desired value of $0.95$ in this case. Further it becomes obvious that the confidence bands obtained by estimating the variance by bootstrap \eqref{bootstrap_var} are always slightly more conservative than their asymptotic versions \eqref{var2} and \eqref{var}, respectively.

 However, considering the bands on $S_1-S_2$ for very small sample sizes, that is $n_1=n_2=20$, the coverage probability lies between $0.91$ and $0.94$ and hence these bands are slightly anti-conservative. The bootstrap bands perform slightly better, but still have coverage probabilites around $0.93$ instead of $0.95$, see the first column of Figure \ref{fig2}. This effect already disappears for $n_1=n_2=50$ where a highly improved accuracy can be observed.
 We observe the same qualitative results for the asymptotic bands for $\log{\tfrac{h_1}{h_2}}$, whereas the bootstrap bands show a  different behaviour, that is being rather conservative for small sample sizes, but also getting more precise with increasing sample sizes.
 
 For smaller sample sizes, all confidence bands under consideration vary in their behaviour over time. This effect gets in particular visible when considering the non-proportional hazards scenario \eqref{scen1b}, see the second row of Figure \ref{fig2}. The coverage probability of the bands for $S_1-S_2$ start with a very accurate approximation during the first two months but then decrease to $0.93$. This effect can be explained by the fact that in the setting of a very small sample, that is $n_1=n_2=20$, after this period of time only very few patients remain (note that the median survival for the reference model is given by $2.6$ months) and hence the uncertainty in estimating the variance becomes rather large. 
 The same holds for all bands under consideration, explaining the decreasing accuracy at later time points.
 
 Finally we consider the scenario of misspecification \eqref{scen2}. Here the hazard ratio varies from $2.5$ to $0.8$ over a time from $1$ to $5$ months, 
 meaning that hazards are non-proportional. The corresponding coverage probabilities are shown in the third row of Figure \ref{fig2}. It turns out that the performance is worse than in case of a correctly identified model as the coverage lies between $0.85$ and $0.9$ and hence below the desired value of $0.95$. 
 However, considering $S_1-S_2$, in a majority of the cases the coverage is above $0.9$, even for a small sample size of $n=40$, where the bootstrap approach performs slightly better than the asymptotic analogue. For increasing sample sizes the coverage, which varies over the time, approximates $0.95$, whereas the bands for  $\log{\tfrac{h_1}{h_2}}$ still do not come sufficiently close to this desired value, even for the largest sample size of $n=200$. Consequently we conclude that these bands suffer more from misspecification than the ones for $S_1-S_2$, where the latter prove to be robust if sample sizes are sufficiently large.


\subsection{Type I errors}
In the following we will investigate type I error rates for the non-inferiority test \eqref{test_diff_ni} and the equivalence test \eqref{test_diff_eq} on the difference of survival curves. 
We consider a nominal level of $\alpha=0.05$ and different sample sizes, i.e.
$(n_1,n_2)=(20,20),\ (n_1,n_2)=(50,50),\ (n_1,n_2)=(100,100)\text{ and }(n_1,n_2)=(150,150)$,
resulting in total sample sizes given by $n=40,100,200$ and $300$, respectively.
As already indicated by the findings concerning the coverage probability presented in Figure \ref{fig2} the difference between asymptotic and bootstrap based bands is very small, in particular for total sample sizes larger than $50$. This becomes also visible in the performance of the test and hence, for the sake of brevity, we only display the results for the asymptotic version here. 

We start by considering scenario \eqref{scen1a} and choose $\theta_2=(1.5,4.9)$, such that the difference curve $S_2(t)-S_1(t)$ attains values of $0.1$, $0.15$ and $0.2$ at time points $1.6$, $2.3$ and $4$, respectively, see Figure \ref{fig1}(a). The median survival is given by $3.8$ months and $2.7$ months, respectively.
In particular we obtain type I errors on the margin of the null hypothesis for every choice of $\delta$. Table \ref{tab1} displays the simulated type I errors, configurations not falling under the null hypothesis are omitted and marked with $"-"$. The first number always corresponds to the equivalence test, the number in brackets displays the type I error of the non-inferiority test, respectively.  It turns out that the approximation of the level is very precise, for the non-inferiority test \eqref{test_diff_ni} in general and for the equivalence test \eqref{test_diff_eq} for sufficiently large sample sizes. In particular for settings on the margin of the null hypothesis, that is $S_2(t_0)-S_1(t_0)=\delta$, we obtain type I errors very close to $0.05$. Only for very small samples, that is $n_1=n_2=20$, the equivalence test \eqref{test_diff_eq} is conservative as the obtained type I errors are zero in almost all scenarios under consideration.

The same arguments hold for the scenario of non-proportional hazards \eqref{scen1b} with $\theta_2=(2,2.5)$, see Table \ref{tab2}. All results obtained here are qualitatively the same as the ones for the scenario above, demonstrating that the presence of a non-constant hazard ratio does not affect the performance of the test.  
Concerning  \eqref{scen2} it turns out that the robustness of the tests with regard to the type I error strongly depends on the equivalence/non-inferiority margin. A type I error inflation occurs for the very liberal margin $\delta=0.2$, whereas for the other choices of $\delta$ the type I error is controlled properly. 

Finally we note that the non-parametric non-inferiority test also yields a precise approximation of the level if sample sizes are sufficiently large. However, the equivalence test is conservative, in particular for $\delta=0.1$. The results and a detailed interpretation are deferred to Section 3.1 of the Supplemental Material.

\begin{table}[h!]
	\centering\small
	\begin{tabular}{c|c|c|c|c}
		$(n_1,n_2)$                  & $(t_0,S_2(t_0)-S_1(t_0))$ & \multicolumn{1}{|c|}{$\delta=0.1$} & \multicolumn{1}{|c|}{$\delta=0.15$} & \multicolumn{1}{|c}{$\delta=0.2$} \\\hline
		\multirow{3}{*}{$(20,20)$}   & $(1.6,0.1)$         &         0.000    (0.049)                     & -                                 & -                                \\
		& $(2.3,0.15)$        &           0.000   (0.025)    &               0.000   (0.051)              & -                                \\
		& $(4,0.2)$           &             0.000   (0.014)      &              0.000  (0.030)           &      0.001 (0.061)                    \\ \hline
		\multirow{3}{*}{$(50,50)$}   & $(1.6,0.1)$         &           0.001 (0.049)                       & -                                 & -                                \\
		& $(2.3,0.15)$        &           0.000 (0.014)                     &          0.017 (0.045)                        & -                                \\
		& $(4,0.2)$           &            0.000 (0.005)                   &         0.010 (0.012)           &       0.047 (0.048)                        \\\hline
		\multirow{3}{*}{$(100,100)$} & $(1.6,0.1)$         &      0.037 (0.037)                         & -                                 & -                                \\
		& $(2.3,0.15)$        &                0.000 (0.009)                 &          0.049   (0.041)                      & -                                \\
		& $(4,0.2)$           &                0.000   (0.001)         &      0.012   (0.006)             &           0.050    (0.051)                \\\hline
		\multirow{3}{*}{$(150,150)$} & $(1.6,0.1)$         &          0.054 (0.044)                 & -                                 & -                                \\
		& $(2.3,0.15)$        &         0.001 (0.002)                       &         0.037  (0.040)       & -                                \\
		& $(4,0.2)$           &             0.000 (0.000)                 &         0.003 (0.003)        &        0.053 (0.048)               \\\hline         
	\end{tabular}
\caption{\small Simulated type I errors of the non-inferiority test \eqref{test_diff_ni} (numbers in brackets) and the equivalence test \eqref{test_diff_eq}  for the scenario of proportional hazards \eqref{scen1a} with $\theta_2=(1.5,4.9)$ at three different time points for different sample sizes and equivalence margins. The nominal level is chosen as $\alpha=0.05$. Scenarios not falling under the null hypothesis are marked with $"-"$.} \label{tab1}
\end{table}

\begin{table}[h!]
	\centering\small
	\begin{tabular}{c|c|c|c|c}
		$(n_1,n_2)$                  & $(t_0,S_1(t_0)-S_2(t_0))$ & \multicolumn{1}{|c|}{$\delta=0.1$} & \multicolumn{1}{|c|}{$\delta=0.15$} & \multicolumn{1}{|c}{$\delta=0.2$} \\\hline
		\multirow{3}{*}{$(20,20)$}   & $(1.9,0.1)$         &     0.000   (0.060)                          & -                                 & -                                \\
		& $(2.4,0.15)$        &              0.000   (0.029)                   &        0.000     (0.049)                        & -                                \\
		& $(3,0.2)$           &                  0.000     (0.016)             &           0.000    (0.021)                   &    0.002     (0.061)                    \\ \hline
		\multirow{3}{*}{$(50,50)$}   & $(1.9,0.1)$         &      0.000     (0.053)        & -                                 & -                                \\
		& $(2.4,0.15)$        &          0.000 (0.011)         &           0.008   (0.052)      & -                                \\
		& $(3,0.2)$           &          0.000    (0.003)       &       0.007       (0.014)      &   0.048  (0.048)                             \\\hline
		\multirow{3}{*}{$(100,100)$} & $(1.9,0.1)$         &       0.002 (0.057)                           & -                                 & -                                \\
		& $(2.4,0.15)$        &      0.000   (0.008)                 &         0.055   (0.055)      & -                                \\
		& $(3,0.2)$           &         0.000 (0.001)               &       0.006 (0.006)      &     0.038   (0.038)           \\\hline
		\multirow{3}{*}{$(150,150)$} & $(1.9,0.1)$         &      0.043  (0.053)         & -                 & -            \\
		& $(2.4,0.15)$        &         0.004    (0.006)         &       0.041  (0.041)         & -                   \\
		& $(3,0.2)$           &           0.000     (0.000)       &      0.005   (0.005)       &    0.049 (0.049)           \\\hline         
	\end{tabular}
	\caption{\small Simulated type I errors of the non-inferiority test \eqref{test_diff_ni} (numbers in brackets) and the equivalence test \eqref{test_diff_eq}  for the scenario of non-proportional hazards \eqref{scen1b} with $\theta_2=(2,2.5)$ at three different time points for different sample sizes and equivalence margins. The nominal level is chosen as $\alpha=0.05$. Scenarios not falling under the null hypothesis are marked with $"-"$.} \label{tab2}
\end{table}

\subsection{Power}
For investigations on the power we consider the same configurations as given above. We now observe the scenario of proportional hazards \eqref{scen1a}  such that the difference curve $S_1(t)-S_2(t)$ attains values of $0.01$, $0.02$ and $0.04$ at time points $0.7$, $1.2$ and $2.3$, respectively. Hence all chosen configurations belong to the alternatives in \eqref{test_diff_ni}  and \eqref{test_diff_eq}.
Table \ref{tab4} displays the simulated power. It turns out that both tests achieve  a reasonable power in most settings under consideration. The power clearly increases with increasing sample sizes and a wider equivalence/non-inferiority margin $\delta$. For instance, when considering $\delta=0.2$ the maximum power is close to or larger than $80\%$ for all sample sizes and both tests. 

For the non-proportional hazards scenario \eqref{scen1b} we consider $\theta_2=(2,3.4)$, resulting in differences of $0.01$ and $0.04$, attained at time points $0.2$ and $0.6$ $(0.01)$ and $3.2$ and $2.7$ $(0.04)$, respectively (see Figure \ref{fig2}). Table \ref{tab5} displays the simulated power. We observe that the power of both tests is reasonably high as well and the results are qualitatively the same as the ones obtained for scenario \eqref{scen1a}. 
Further it becomes obvious that for later time points but equal distances between the two survival curves the power decreases, in particular in presence of small sample sizes. This can be explained by the fact that the remaining subjects become less with progressing time, resulting in a higher uncertainty after 3 months compared to $0.2$ and $0.6$ months, respectively. Of note, at $3$ months more than half of the subjects experienced an event in this scenario.

In summary we observe a reasonably high power for both tests.
The power increases with increasing sample sizes and a wider equivalence margin. Further the presence of non-proportional hazards does not affect the power, meaning that in both scenarios the results are qualitatively the same. 
Finally we also observe that for the non-parametric approach the power is much smaller than for our new method, in particular for small sample sizes and for the scenario of non-proportional hazards, which underlines the theoretical findings. Detailed results can be found in Section 3.1 of the Supplemental Material.

\begin{table}[h!]
	\centering\small
	\begin{tabular}{c|c|c|c|c}
		$(n_1,n_2)$                  & $(t_0,S_2(t_0)-S_1(t_0))$ & \multicolumn{1}{|c|}{$\delta=0.1$} & \multicolumn{1}{|c|}{$\delta=0.15$} & \multicolumn{1}{|c}{$\delta=0.2$} \\\hline
		\multirow{3}{*}{$(20,20)$}   & $(0.7,0.01)$         &      0.145  (0.436)          &     0.512  (0.723)        &              0.795  (0.880)               \\
		& $(1.2,0.02)$        &       0.002   (0.227)    &       0.057   (0.410)       &         0.303   (0.595)              \\
		& $(2.3,0.04)$           &          0.000 (0.114)     &        0.000 (0.208)       &        0.000   (0.365)                  \\ \hline
		\multirow{3}{*}{$(50,50)$}   & $(0.7,0.01)$         &     0.553  (0.696)         &     0.929  (0.943)           &       0.995       (0.995)                 \\
		& $(1.2,0.02)$        &        0.025  (0.360)          &         0.501  (0.645)        &         0.833   (0.870)                  \\
		& $(2.3,0.04)$           &       0.000  (0.187)       &       0.085  (0.335)       &       0.460    (0.601)                 \\\hline
		\multirow{3}{*}{$(100,100)$} & $(0.7,0.01)$         &       0.887 (0.906)       &       0.998 (0.998)      &   1.000     (1.000)                   \\
		& $(1.2,0.02)$      &       0.388  (0.548)       &      0.858  (0.875)    &        0.988   (0.998)        \\
		& $(2.3,0.04)$           &          0.017  (0.274)        &       0.486  (0.557)        &         0.854  (0.861)            \\\hline
		\multirow{3}{*}{$(150,150)$} & $(0.7,0.01)$         &     0.973  (0.974)         &       1.000  (1.000)       &     1.000   (1.000)       \\
		& $(1.2,0.02)$       &         0.639 (0.703)     &        0.971 (0.971)      &         0.999   (0.999)        \\
		& $(2.3,0.04)$           &          0.236   (0.364)      &      0.728   (0.736)        &          0.949   (0.950)           \\\hline         
	\end{tabular}
	\caption{\small Simulated power of the non-inferiority test \eqref{test_diff_ni} (numbers in brackets) and the equivalence test \eqref{test_diff_eq}  for the scenario of proportional hazards \eqref{scen1a} with $\theta_2=(1.5,3.7)$ at three different time points for different sample sizes and equivalence margins. The nominal level is chosen as $\alpha=0.05$. } \label{tab4}
\end{table}

\begin{table}[h!]
	\centering\small
	\begin{tabular}{c|c|c|c|c}
		$(n_1,n_2)$                  & $(t_0,S_2(t_0)-S_1(t_0))$ & \multicolumn{1}{|c|}{$\delta=0.1$} & \multicolumn{1}{|c|}{$\delta=0.15$} & \multicolumn{1}{|c}{$\delta=0.2$} \\\hline
		\multirow{4}{*}{$(20,20)$}   & $(0.2,0.01)$         &   0.964 (0.964)   &     0.996   (0.999)           &      1.000   (1.000)           \\
		& $(0.6,0.04)$        &    0.372  (0.448)           &     0.712  (0.712)       &       0.891  (0.893)           \\
		& $(2.7,0.04)$        &      0.000 (0.124)    &       0.000   (0.219)     &    0.000  (0.297)              \\
		& $(3.2,0.01)$           &         0.000  (0.179)    &      0.000  (0.284)     &       0.000    (0.374)                      \\ \hline
		\multirow{4}{*}{$(50,50)$}   & $(0.2,0.01)$         &     1.000 (1.000)      &     1.000  (1.000)         &    1.000   (1.000)           \\
		& $(0.6,0.04)$        &        0.637  (0.640)     &     0.938  (0.938)        &     0.997  (0.997)             \\
		& $(2.7,0.04)$        &        0.000 (0.151)    &          0.047 (0.339)     &      0.436  (0.563)            \\
		& $(3.2,0.01)$           &           0.000  (0.258)    &     0.054  (0.468)      &   0.481   (0.684)                           \\ \hline
		\multirow{4}{*}{$(100,100)$}   & $(0.2,0.01)$         &         1.000  (1.000)               &    1.000   (1.000)            &              1.000    (1.000)            \\
		& $(0.6,0.04)$        &       0.818  (0.841)      &     0.995  (0.995)      &    1.000    (1.000)            \\
		& $(2.7,0.04)$        &       0.001 (0.246)  &      0.442 (0.555)     &  0.803   (0.833)               \\
		& $(3.2,0.01)$           &       0.006  (0.406)    &      0.558  (0.728)        &            0.869  (0.921)                   \\ \hline
	\multirow{4}{*}{$(150,150)$}   & $(0.2,0.01)$    &      1.000 (1.000)        &        1.000  (1.000)          &               1.000  (1.000)               \\
	& $(0.6,0.04)$        &     0.927 (0.927)       &         1.000  (1.000)    &        1.000   (1.000)          \\
	& $(2.7,0.04)$        &        0.206  (0.326)     &       0.678 (0.701)    &     0.922  (0.928)             \\
	& $(3.2,0.01)$           &      0.267  (0.553)    &       0.797  (0.859)        &      0.903   (0.984)                        \\ \hline    
	\end{tabular}
	\caption{\small Simulated power of the non-inferiority test \eqref{test_diff_ni} (numbers in brackets) and the equivalence test \eqref{test_diff_eq}   for the scenario of non-proportional hazards \eqref{scen1b} with $\theta_2=(2,3.4)$ at four different time points for different sample sizes and equivalence margins. The nominal level is chosen as $\alpha=0.05$. } \label{tab5}
\end{table}

\section{Case Study}
\label{sec:data}
\def\theequation{4.\arabic{equation}}

In the following we investigate a well known benchmark dataset regarding survival analysis. The data set veteran from VeteranÆs Administration Lung Cancer Trial, published in \cite{kalbfleisch1983} and implemented in the R package survival by \cite{survival-package}, describes a two-treatment, randomized trial for lung cancer.
In this trial, male patients with advanced inoperable lung cancer were allocated to either a standard therapy (reference treatment, $\ell=1$) or a chemotherapy (test treatment, $\ell=2$). Numerous covariates were documented, 
including time to death for each patient, which is the primary endpoint of our analysis. In total 137 observations, allocated to $n_1=69$ patients in the reference group and $n_2=68$ in the test group, are given. 

The R code reproducing the results presented in the following can be found online at \url{https://github.com/kathrinmoellenhoff/survival}. As our analysis is model-based, we start with a model selection step. More precisely we split the data into the reference group and the test group and assume six different distributions, that is a Weibull distribution, an exponential distribution, a Gaussian distribution, a logistic distribution, a log-normal distribution and a log-logistic distribution, respectively. We fit the corresponding models separately per treatment group, resulting in 12 models in total.  Finally we compare for each group the six different models using Akaike's Information Criterion (AIC, see \cite{sakamoto1986}). 
It turns out that for the group receiving the reference treatment the Weibull and the exponential model provide the best fits (AICs are given by $749.1$, $747.1$, $799.9$, $794.7$, $755.1$ and $758.1$ in the order of the models mentioned above) whereas in the test group the log-logistic, the log-normal and the Weibull model are the best ones (AICs given by $749.1$, $750.1$ and $751.7$, respectively). Therefore we decide to base our analyses on Weibull models for both groups. However, note that all tests could also be performed assuming  different distributions for each treatment. 

\begin{figure}
	\scriptsize
	\begin{center}
		(a)\subfigure{\includegraphics[width=0.46\textwidth]{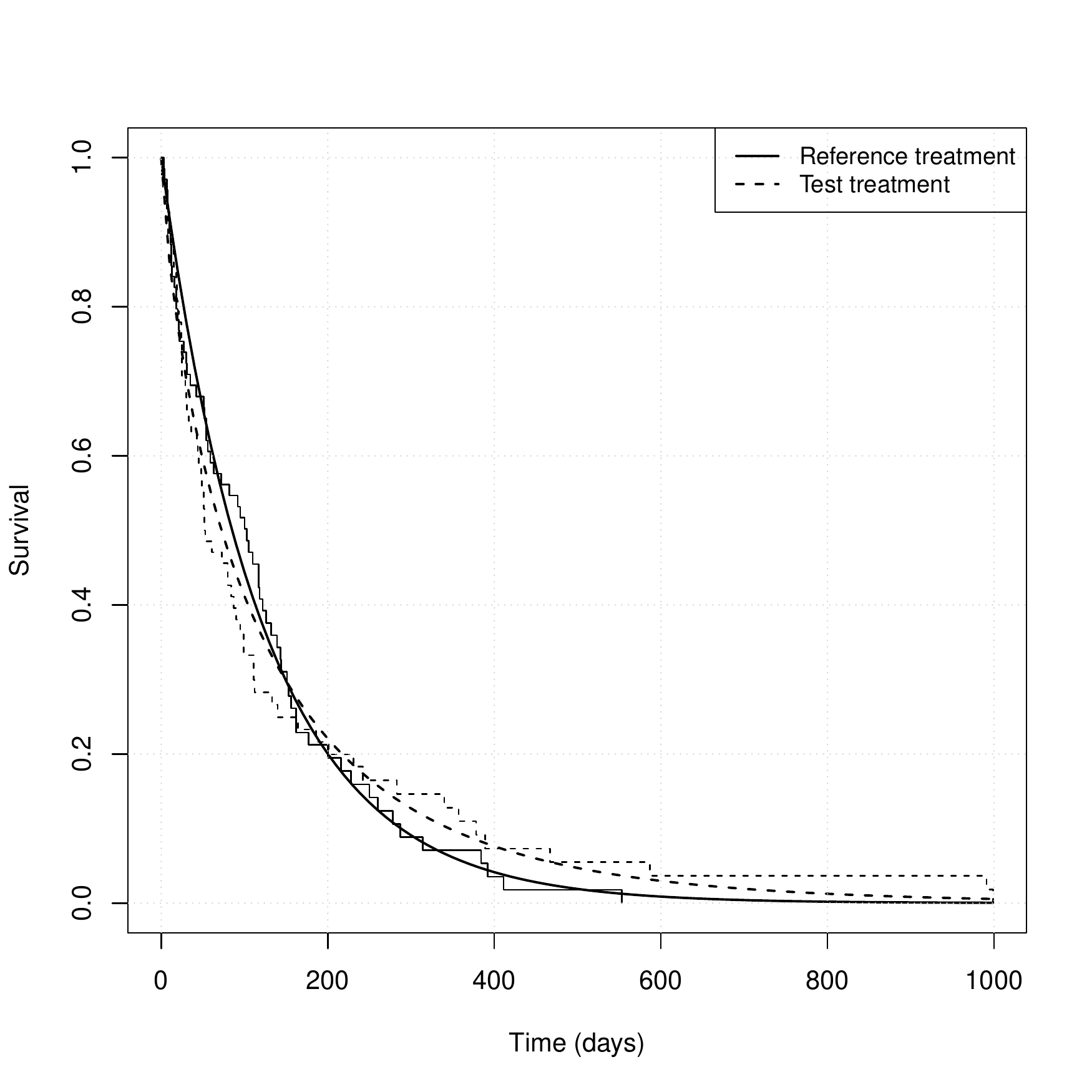}}
		(b)\subfigure{\includegraphics[width=0.46\textwidth]{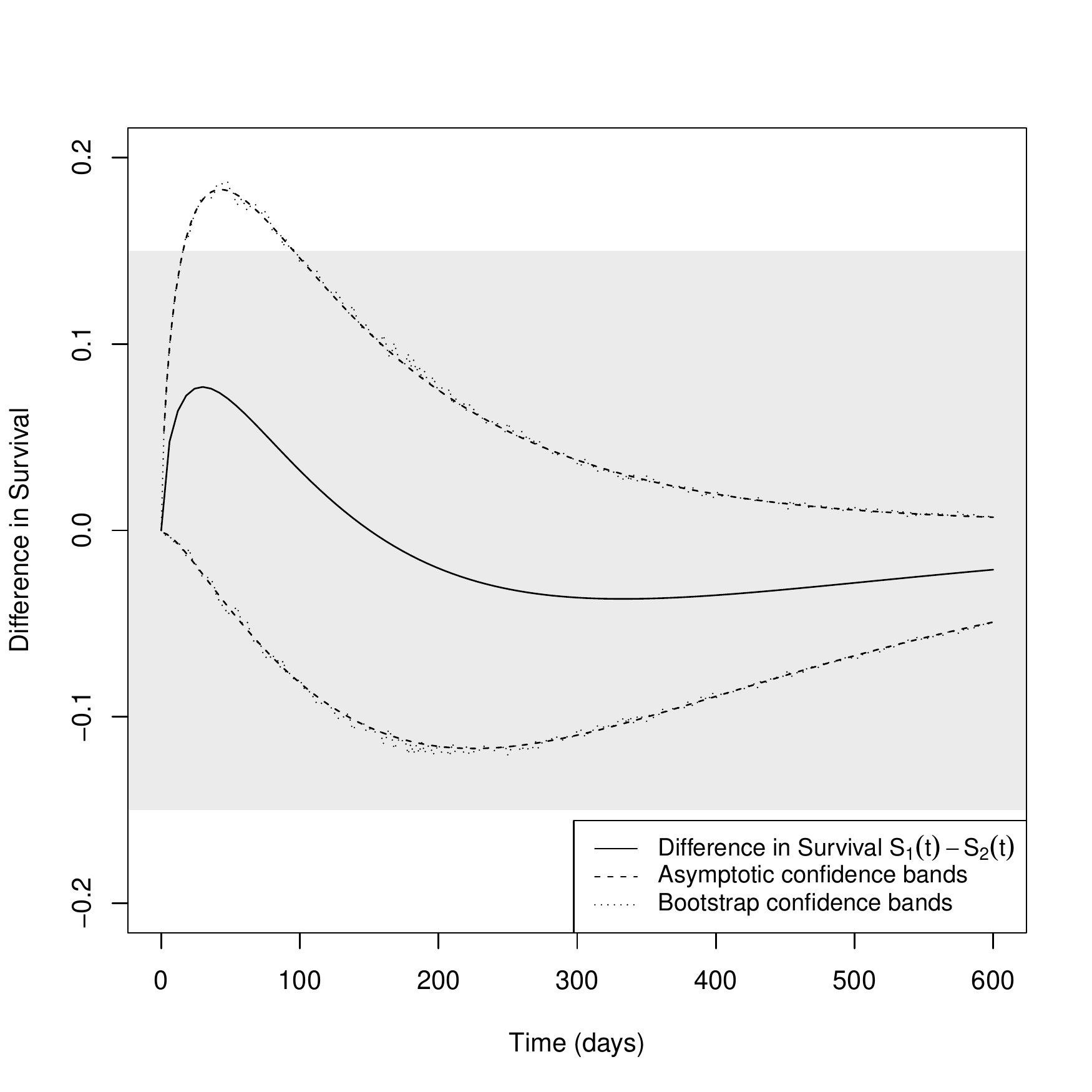}}
		\caption{\small (a): Survival curves (Kaplan-Meier curves and Weibull models) for the veteran data. Solid lines correspond to the reference group, dashed lines to the test treatment.  (b): Difference in survival, pointwise confidence bands obtained by the asymptotic approach (dashed) and bootstrap (dotted), respectively, on the interval $\left[ 40,600\right] $. The shaded area indicates the equivalence margins with $\delta=0.15$.}\label{fig1cs}\end{center}
\end{figure}

For the analysis we now consider the complete likelihood function \eqref{mle} incorporating the distribution of the censoring times. We assume the censoring times to be exponentially distributed and maximizing the likelihood \eqref{mle} yields 
$ \hat\theta_1=(4.82,1.01),\ \hat\psi_1=0.00063$ and $ \hat\theta_2=(4.76,1.3),\ \hat\psi_2=0.00046.$
Figure \ref{fig1cs}(a) displays the corresponding Weibull models and the non-parametric analogue given by Kaplan-Meier curves. It turns out that for both treatment groups the parametric and the non-parametric curves are very close to each other. Further we observe that the survival curves of the two treatment groups cross each  other which indicates that the assumption of proportional hazards is not justified here. Indeed, the hazard ratio ranges from $0.55$ to $1.93$ from the first time of event (3 days) until the end of the observational period (999 days) and therefore an analysis using a proportional hazards model is actually not applicable here. 
The p-value of the log-rank test is $0.928$ and thus does not detect any difference between the two groups.

We will now perform a similar analysis using the parametric models and the theory derived in Section \ref{sec2}. For the sake of brevity we will only consider difference in survival, analyses concerning the (log) hazard ratio can be conducted in the same manner. We consider the first 600 days of the trial. We set $\alpha=0.05$ and  calculate lower and upper $(1-\alpha)$-pointwise confidence bands according to \eqref{conf_sf} at several points $0\leq t \leq 600$. Estimates of the variance $\hat\sigma_{\Delta}$ are obtained by both, the asymptotic approach and bootstrap as described in Algorithm \ref{alg1}, respectively. 
Figure \ref{fig1cs}(b) displays the estimated difference curve $\Delta(t,\hat\theta_1,\hat\theta_2)=S_1(t,\hat\theta_1)-S_2(t,\hat\theta_2)$ and the pointwise confidence bands on the interval $\left[ 0,600\right] $. It becomes obvious that there is almost no difference between the two confidence bands, meaning that the asymptotic and the bootstrap approach yield very similar results here, which can be explained by the rather high sample size combined with the very low rate of censoring.

We start our analysis considering $t_0=80$, which is close to the median survival of both treatment groups. The difference in survival is $\Delta(80,\hat\theta_1,\hat\theta_2)=0.047$  and the asymptotic confidence interval is given by $\left[-0.068,0.163\right]$, while the bootstrap yields $\left[-0.067,  0.162\right]$. Note that these are  two-sided $90\%$-confidence intervals, as we use $95\%$-upper and $95\%$-lower confidence bands for the test decisions.
Investigating the hypotheses \eqref{hypotheses_ni} and \eqref{hypotheses_eq} we observe that both, non-inferiority and equivalence can be claimed for all $\delta>0.163$ $(0.162)$, respectively. Consequently, for $\delta=0.15$, which is indicated by the shaded area  in Figure \ref{fig1cs}(b), $H_0$ cannot be rejected in both cases, meaning in particular that the treatments cannot be considered equivalent with respect to the survival at day $80$.
Figure \ref{fig1cs}(b) further displays these investigations for $\delta=0.15$ simultaneously at all time points under consideration. We conclude that the chemotherapy is non-inferior to the standard therapy after 96 days, as this is the first time point where the upper confidence bound is smaller than $\delta=0.15$. The same conclusion can be made concerning equivalence as for all t in $\left[ 0,600\right]$, the lower confidence bounds are completely contained in the rejection region, meaning that they are larger than $-\delta=-0.15$. 
Concluding we observe that considering for instance $\delta=0.2$, equivalence would be claimed at all time points under consideration, demonstrating the impact of the choice of the margin. 

\section{Conclusions}
\label{sec:conc}

In this paper we addressed the problem of survival analysis in presence of non-proportional hazards. Here,  commonly used methods, as Cox's proportional hazards model or the log-rank test, are not optimal and suffer from a loss of power.
Therefore we proposed another approach for investigating equivalence or non-inferiority of time-to-event outcomes based on the construction of (pointwise) confidence intervals and applicable irrespectively of the shape of the hazard ratio. 
We constructed confidence bands for both, the (log) hazard ratio and the difference of the survival curve proposing two alternatives, one based on asymptotic investigations and the other on a parametric bootstrap procedure. Both approaches show a similar performance, where the latter has the advantage that it can be used for very small samples as well and without the need to calculate the asymptotic variance. 

Our approach provides a framework for investigating survival in many ways. Apart from specific time points entire periods of time can be observed. 
We demonstrated that the presence of non-proportional hazards has no effect on the performance of the confidence bands and the non-inferiority and equivalence test, respectively, which means that they do not depend on this assumption, providing a much more flexible tool compared to standard methodology. 

Our methods are based on the estimation of  parametric survival functions, which can be a very powerful tool once the suitability of the model has been proven.  This has to be assessed in a preliminary study, using for instance model selection criteria as the AIC.  
We demonstrated the robustness of our approach even in the case of misspecification.
An interesting extension of the methodology could be given by model averaging which possibly further improves the performance of the procedures. 



\bibliography{survival_bib}
\end{document}